\documentclass[12pt]{article}

\usepackage{graphicx}% Include figure files

\newcommand{\beq}{\begin{equation}}
\newcommand{\eeq}{\end{equation}}
\newcommand{\beqa}{\begin{eqnarray}}
\newcommand{\eeqa}{\end{eqnarray}}
\newcommand{\beqar}{\begin{eqnarray*}}
\newcommand{\eeqar}{\end{eqnarray*}}

\begin{document}
\thispagestyle{empty}

%\hfill{\sc UG-FT-276/10}
%\vspace*{-2mm}
%\hfill{\sc CAFPE-146/10}

\vspace{32pt}
\begin{center}
{\textbf{\Large Propagation in the atmosphere}}

{\textbf{\Large of ultrahigh-energy charmed hadrons}}

\vspace{40pt}

R.~Barcel\'o, J.I.~Illana, M.D.~Jenkins, M.~Masip
\vspace{12pt}

\textit{
CAFPE and Departamento de F{\'\i}sica Te\'orica y del
Cosmos}\\ \textit{Universidad de Granada, E-18071, Granada, Spain}\\
\vspace{16pt}
\texttt{rbarcelo@ugr.es, jillana@ugr.es, mjenk@ugr.es, masip@ugr.es}
\end{center}

\vspace{40pt}

\date{\today}% It is always \today, today,

\begin{abstract}

Charmed mesons may be produced when a primary cosmic ray or the
leading hadron in an air shower collide with an atmospheric
nucleon. At energies $\ge 10^8$ GeV their decay length becomes
larger than 10 km, which implies that they tend to interact in the
air instead of decaying. We study the collisions of long-lived
charmed hadrons in the atmosphere. We show that
$(\Lambda_c,D)$--proton diffractive processes and partonic
collisions of any $q^2$ where the charm quark is an spectator have
lower inelasticity than $(p,\pi)$--proton collisions. In particular,
we find that a $D$ meson
deposits in each interaction 
just around 55\% of the energy deposited by a pion. 
On the other hand, collisions involving the valence $c$ quark
(its annihilation with a sea $\bar c$ quark in the target 
or $c$-quark exchange in the $t$ channel) may deposit most 
of the $D$ meson
energy, but their frequency is low (below 0.1\% of inelastic
interactions). As a consequence, very energetic 
charmed hadrons may keep a significant fraction of their initial 
energy after several hadronic interactions, reaching
much deeper in the atmosphere than pions or protons
of similar energy.

%\pacs{13.85.-t,96.50.sd,14.40.Lb}% PACS, the Physics and Astronomy
%\keywords{Extensive air showers, charmed hadrons}
                              %Use showkeys class option if keyword
\end{abstract}

\newpage

\section{Introduction}

Cosmic rays reach the Earth with energies of up to
$10^{11}$ GeV. When they enter the atmosphere they experience
ultrahigh energy collisions, probing a scale not accessible
at colliders. These processes are interesting because
they could produce exotic particles or involve interactions
not seen at lower energies.
There is, however, a second generic reason that make these
cosmic-ray collisions interesting:
they probe a regime where the properties of the standard
particles may be substantially {\it different}.

In particular, here we will focus on the charm quark. 
Charm-quark
pairs may be abundant inside extensive air showers
\cite{Brodsky:1980pb,Berghaus:2007hp}. Once produced,
each quark will result into a 
$D^{+,0}$ meson or a $\Lambda_c^+$ baryon,
that will then decay weakly into light hadrons and (with a
$\approx 0.1$ branching ratio) leptons. These processes have been
extensively studied in the literature, as they may be the dominant
source of atmospheric muons and neutrinos of energy $E\ge 10^6$
GeV \cite{Gondolo:1995fq,Costa:2000jw,Illana:2009qv,Illana:2010gh}. 
A {\it new} regime, however, would be achieved if the charmed
hadron has an energy $E>E_c$ such that its decay length
($\lambda_{d}=c\tau E/m$) becomes larger than its interaction
length ($\lambda_i$) in the atmosphere. For example,
a  $10^8$ GeV $D^+$ meson is effectively a {\it long-lived}
($\lambda_d^{D^+}\approx 18$ km) heavy hadron that interacts and
propagates in the air instead of decaying. Such a behaviour,
{\it unknown of} in colliders (where a $D$ meson decays before it
has experienced any type of interaction) could look somewhat
similar to the one of the $R$--hadrons predicted in some 
supersymmetric models \cite{Baer:1998pg,Illana:2006xg}. 
The propagation in matter of a charmed hadron is of no
interest at the Tevatron or the LHC, but it is most likely
important in extensive air-shower experiments.

In this article we analyze the collisions of very energetic
$D$ mesons and
$\Lambda_c^+$ baryons
with protons at rest. Our objective is
to estimate the amount of energy that they deposit
in each hadronic interaction in the atmosphere. First we study
diffractive processes, using the Montecarlo  
code PYTHIA \cite{Sjostrand:2006za} to
simulate the collisions. Then we consider partonic collisions
of any $q^2$, with our analysis also based on PYTHIA. 
Finally we estimate the effects of going from
a proton to a nucleus target.

\section{Diffractive processes}

A charmed hadron $H$ is basically different from a pion or a
proton in the fact that it contains a heavy {\it core} of mass
$m_c\approx 1.27$ GeV. 
If one thinks of a proton as three {\it clouds} of mass 
$m=0.3$ GeV associated to the three constituent quarks, then 
the $\Lambda_c$ baryon consists also of three similar clouds,  
but one of them has an additional electroweak core 
and a total mass $m_c+m=1.6$ GeV.
In a long-distance (low-$q^2$)
collision with an atmospheric nucleon, however, this heavy core
will be {\it invisible} (the proton and the $\Lambda_c$ clouds 
will look identical), 
as only interactions of $q^2\ge m_c^2$
can resolve it. Therefore, we will assume that the momentum
exchanged through pomerons or other non-perturbative dynamics with
the target does not depend on the electroweak core in $H$.

Let us be more specific. We separate the mass
$m_H$ of the heavy hadron into
\beq
m_H=m_c+\Lambda
\;.
\label{mH}
\eeq
The light degrees of freedom in $H$ carry just
a fraction $w\equiv (m_H-m_c)/m_H$ of the hadron energy $E$, so
in a diffractive process $H$ will be {\it seen} by the target
nucleon like a light hadron of energy $wE$. Therefore,
to estimate the momentum $q^\mu$ absorbed by $H$ in the
process, we will just simulate with PYTHIA
the collision of a proton (for $H=\Lambda_c$) or a pion
(for $H=D$) of energy $wE$ with the nucleon and will
assume that $q^\mu$ is the same when the incident particle
is the charmed hadron.

Once $H$ absorbs the momentum $q^\mu$ it becomes a {\it
diffractive} system $H_{di\! f}$ of mass $M$, the critical
parameter in the collision. If $M< m_H+1$ GeV the collision is
{\it quasielastic} and $H_{di\! f}$ will just decay into two
bodies ({\it e.g.}, $H+\eta$). For larger values of $M$ the system
is treated by PYTHIA \cite{Navin:2010kk}
like a string with the quantum numbers of
$H$. When $H$ is a baryon the string may be stretched 
between a quark and a diquark or between a 
quark, a gluon and a diquark, whereas for a diffractive meson
the string connects a quark and an antiquark or a quark, a gluon 
and an antiquark.

To illustrate our procedure, let us consider the diffractive
collision of a $10^9$ GeV $\Lambda_c^+$ baryon
($m_{\Lambda_c}=2.28$ GeV) with a proton ($m_{p}=0.94$ GeV) at
rest. We need to simulate the collision of a $4.43\times 10^8$ GeV
proton, and we will do it in the c.o.m. frame, where each proton
carries 14.4 TeV.

\begin{enumerate}

\item[{\it (i)}] In a first PYTHIA  example the diffracted proton
absorbs $q^\mu=(-1.3;-0.13,0.10,-1.3)$ GeV, getting a mass of
$M=1.58$ GeV. The system then decays into a $\Delta^+ \pi^0$, and
the $\Delta^+$ finally produces a proton and another $\pi^0$.
Going back to the lab frame we find the final proton with
$E'_p=2.75\times 10^8$ GeV, {\it i.e.}, the leading baryon carries
a fraction $0.62$ of the initial energy. The elasticity $z$ in 
this collision is then $z=0.62$. If the incident
particle is a $10^9$ GeV $\Lambda_c^+$, we first go to the same
$pp$ c.o.m. frame (where the energies of the charmed baryon
and the target proton are 32.5 TeV and 14.4 TeV, respectively).
The $\Lambda_c^+$ absorbs there the same momentum $q^\mu$ and gets
a diffractive mass of $M=3.2$ GeV, it then goes into
$\Sigma_c^+\pi^0$, and the $\Sigma_c$ decays into a $\Lambda_c^+$
and another $\pi^0$. Going back to the lab frame we obtain that
the final $\Lambda_c^+$ baryon carries $E'=9.3\times 10^8$ GeV,
implying an elasticity $z=0.93$. This type of quasielastic
processes accounts for $8\%$ of all diffractive collisions.

\item[{\it (ii)}] In a second PYTHIA simulation the momentum
absorbed by the incident proton in the c.o.m. frame is
$q^\mu=(0.12;0.14,0.14,-0.12)$ GeV, which produces a diffractive mass
$M=83$ GeV. The system evolves into a string stretching between a
$u$-quark and a $(u d)$-diquark. Quark fragmentation and the
(strong or electromagnetic) 
decay of baryonic resonances results then into a
leading baryon plus 18 other hadrons of lower energies. In
particular, in the lab frame there is a $p$ of $1.8\times 10^8$
GeV ({\it i.e.}, $z=0.41$) plus 16 mesons (pions and kaons) and a
$p\, \bar n$ pair sharing the rest of the energy. Changing the proton
for a $10^9$ GeV $\Lambda_c^+$ we obtain a diffractive mass of
$M=124$ GeV. Now the system may define 3 different (equally
probable) diquark--quark strings: $(c d)$--$u$; $(c u)$--$d$ or
$(u d)$--$c$. The first two cases tend to result into a leading
charmed baryon after the collision. For example, in the first case
we obtain a final $\Lambda_c^+$ of $6.0\times 10^8$ GeV
($z=0.60$). The third case, however, is basically different, as
the struck quark is the charm and it will most likely fragment
into a $D$ meson. With the PYTHIA simulation we obtain a final
$D^+$ of $3.8\times 10^8$ GeV ($z=0.38$). In around 30\% of
diffractive collisions the $\Lambda_c$ {\it changes} into a $D$
meson. In contrast, in a diffractive proton--proton collision the
leading hadron becomes a meson in just 15\% of the cases.

\item[{\it (iii)}] In the final example PYTHIA provides an 
event with large diffractive mass ($M=6087$ GeV;
$q^\mu=(641;0.36,0.43,-641)$ GeV) and a diquark--gluon--quark
string. The final spectrum is similar to the one in case {\it
(ii)}, although with a larger multiplicity of final states and a
lower energy in the leading hadron. Whereas in the $pp$ collision
we find a neutron carrying a $38\%$ of the initial energy, in the
$\Lambda_c p$ process we obtain a $4.1\times 10^8$ GeV ($z=0.41$)
$\Lambda_c^+$ or a $7.3\times 10^8$ GeV ($z=0.73$) $D^0$ depending
on the flavor of the struck quark.

\end{enumerate}

\begin{figure}
\begin{center}
\includegraphics[width=0.55\linewidth]{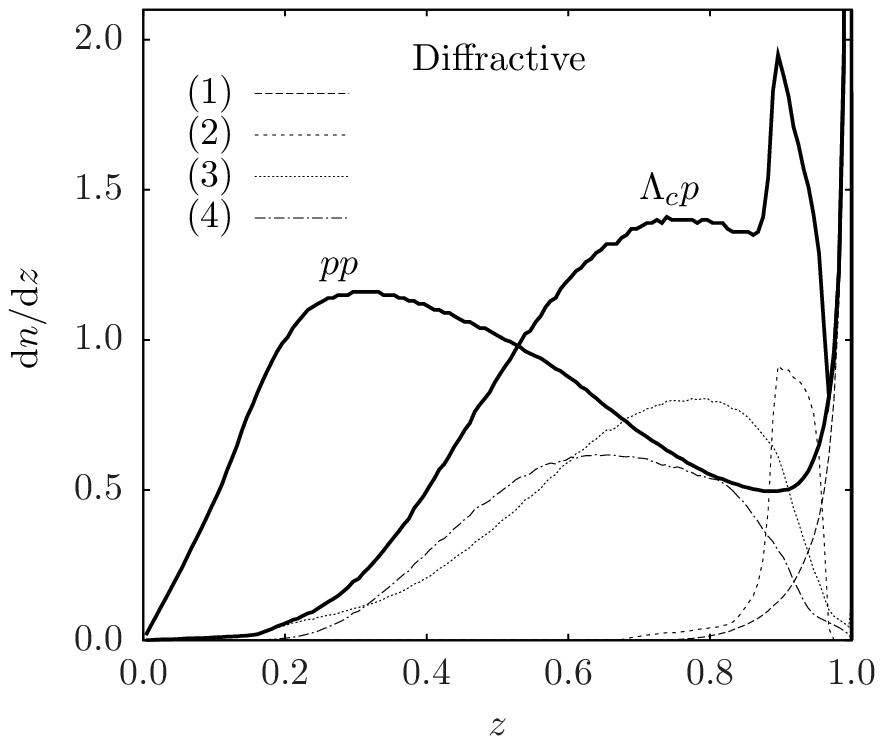} 
\end{center}
\caption{Fraction $z$ of energy taken by the leading baryon
or the charmed hadron in $p p$ and $\Lambda_c p$ 
diffractive collisions,
respectively. We have separated the distributions in the
4 types of $\Lambda_c p$ processes described in the text.
\label{fig1}}
\end{figure}

In Fig.~1 we plot the distribution of the fraction $z$ of energy
taken by the leading baryon after a $pp$ diffractive collision
or by the charmed hadron after the analogous $\Lambda_c p$
process. The average values of the distributions are $\langle
z\rangle =0.60$ and $\langle z\rangle =0.77$, respectively, which
imply an inelasticity $K=1-\langle z\rangle$:
\beq 
K_{pp}^{di\!
f}=0.40\;;\;\;\;\;K_{\Lambda_c p}^{di\! f}=0.23\,. 
\eeq 
The regions of highest $z$ (with $K_{\Lambda_c p}^{(1)} = 0.01$)
correspond to 28\% of diffractive processes where the incident
hadron does not break (but the target proton does), whereas 
8\% of  processes correspond to quasielastic events with 
$M<m_{\Lambda_c}+1$ GeV and have a
slightly higher inelasticity, 
$K_{\Lambda_c p}^{(2)} =0.09\pm 0.04$, where
$\Delta K=\sqrt{\langle z^2 \rangle -
\langle z \rangle^2}$. 
In the remaining collisions the $\Lambda_c$ may become a $D$ meson
or a charmed baryon (29\% and 35\% of all diffractive processes,
respectively). The $D$ mesons carry a fraction $\langle
z\rangle=0.64$ of the initial energy ({\it i.e.}, 
$K_{\Lambda_c p}^{(3)} =0.36\pm 0.16$), whereas charmed baryons imply a 
slightly lower
inelasticity, $K_{\Lambda_c p}^{(4)} =0.32\pm 0.17$. The leading $D$ meson
may be a $D^+$, a $D^0$ or a $D_s$ in an approximate proportion of
$1 : 3 : 0.5$ (the $D^+$--$D^0$ isospin symmetry is broken by the
decay of charm resonances), whereas the baryons are mostly
$\Lambda_c^+$ (94\%) with some $\Xi_c^0$ (3\%) and $\Xi_c^+$ (3\%).
In Fig.~1 we have separated these four types of diffractive
$\Lambda_c p$ processes.

\begin{figure}
\begin{center}
\includegraphics[width=0.55\linewidth]{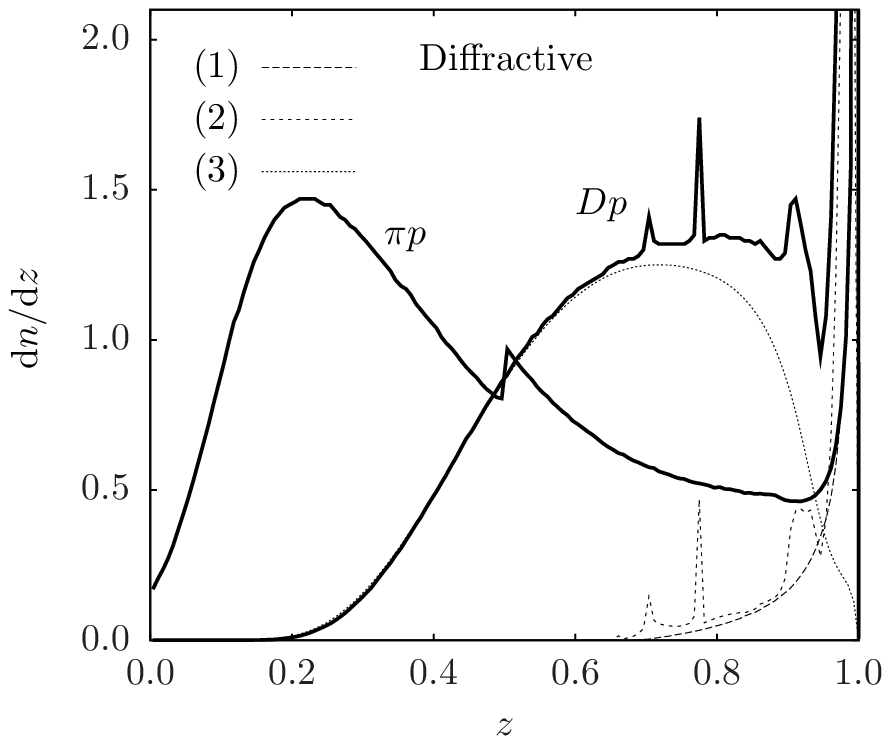} 
\end{center}
\caption{Fraction $z$ of energy taken by the leading baryon
or the charmed hadron in $\pi p$ and $D p$ 
diffractive collisions,
respectively. We have separated the distributions in the
3 types of $D p$ processes described in the text.
\label{fig2}}
\end{figure}

For an incident $10^9$ GeV $D^+$ meson we proceed in an analogous
way, studying the diffractive collision of a $3.2\times 10^8$ GeV
pion with a proton at rest. We simulate with PYTHIA the collision
in the c.o.m. frame, we read the $q^\mu$ absorbed by the pion and
assume that the momentum absorbed by a $D^+$ in that frame would
be the same. We find that in 24\% of the processes the incident
meson remains unbroken ({\it i.e.}, $M=m_{D}$), whereas
quasielastic events ($M<m_{D}+1$ GeV) account for 15\% of the
total. The inelasticity of these two types of events is $K_{\pi
p}^{(1)}= 0.02$ and $K_{\pi p}^{(2)}=0.05\pm 0.05$, 
respectively. The rest
(61\%) of diffractive collisions deposit an average fraction
$K_{\pi p}^{(3)}=0.33\pm 0.17$ of the initial energy. In Fig.~2 we give
the $z$ distribution for the three types of events. The {\it peaks}
in $D$ collisions appear at $z\approx m_D/(m_D+m_{\eta,\omega},...)$,
whereas the {\it step} in the pion distribution results 
from processes
with two final pions of similar energy. 
The average 
inelasticity is
\beq
K_{\pi p}^{di\! f}=0.47\;;\;\;\;\;K_{D
p}^{di\! f}=0.21\;.
\eeq
Whereas a pion loses energy faster than a
proton, the diffractive collisions of a $D$-meson are 
slightly less
inelastic than the ones of a charmed baryon.

Finally, the diffractive cross section for a $10^9$ GeV charmed
hadron can be read from the PYTHIA estimate for a $4.4\times 10^8$
GeV proton or a $3.2\times 10^8$ GeV pion, 
\beq
\sigma_{di\! f}^{\Lambda_c p}=26.2\; {\rm mb}\;; \;\;\;
\sigma_{di\! f}^{D p}=15.6 \; {\rm mb}\;. 
\eeq

\section{Partonic collisions}

Non-diffractive processes dominate the inelastic cross section in
PYTHIA simulations. In particular, at $4.43\times 10^8$ GeV the
cross section for parton-parton interactions of any $q^2$ in
$p p$ and $\pi p$ collisions is\footnote{PYTHIA extrapolates
to $q^2<2$ GeV using a minimum-biased method.}
\beq
\sigma_{n-di\! f}^{p p}=62.2\; {\rm mb}\;; \;\;\;\sigma_{n-di\!
f}^{\pi p}=43.6\; {\rm mb}\,.
\eeq
We will use these simulations
to obtain an approximate description of partonic $H p$
collisions.

We model $H$ as a system with the same parton content as the
corresponding proton or pion but substituting a valence up quark
$u_0$ for the charm quark. Like in diffractive processes, we will
associate a charmed hadron $H$ of energy $E$ to a light hadron of
energy $E(m_H-m_c)/m_H$. If $u_0$ carries a fraction $x$ of the
proton or pion momentum we will change it for a $c$ with
\beq
x_c={m_c\over m_H} + {m_H-m_c\over m_H} x\,.
\eeq
In this way the
excess of energy in $H$ is carried entirely by the charm quark,
whereas the light partons in both hadrons ($H$ and $p$ or $\pi$)
carry exactly the same amount of energy.

We will then distinguish two types of $H p$ non-diffractive
collisions: those where the charm is an spectator ({\it i.e.}, it
is a light parton in $H$ who hits a parton in the target proton),
and processes where the charm itself interacts. For the first one
we will just simulate with PYTHIA the parton process using a light
hadron and then {\it substitute} the spectator $u_0$  for the
charm quark. Charm interactions, on the other hand, have a much
smaller cross section, 
\beq 
\sigma_{c\;int}^{H p}=0.8 \; {\rm mb}\,, 
\eeq 
than the processes with an spectator charm quark, 
\beq
\sigma_{c\;spec}^{\Lambda_c p}=61.4 \;{\rm mb}\,; \;\;\;
\sigma_{c\; spec}^{D p}=42.8\; {\rm mb}\,, 
\eeq 
but they imply
collisions of higher inelasticity. In particular, there is the
possibility that the $c$ quark in the incident $H$ hadron
annihilates with a sea $\bar c$ in the target proton. The charmed
hadron after such process has lost basically {\it all} its energy.
These events, however, occur in just a fraction of all partonic
processes, \beq \sigma_{c\bar c}^{H p}=0.02\; {\rm mb}\,. \eeq

\begin{figure}
\begin{center}
\includegraphics[width=0.55\linewidth]{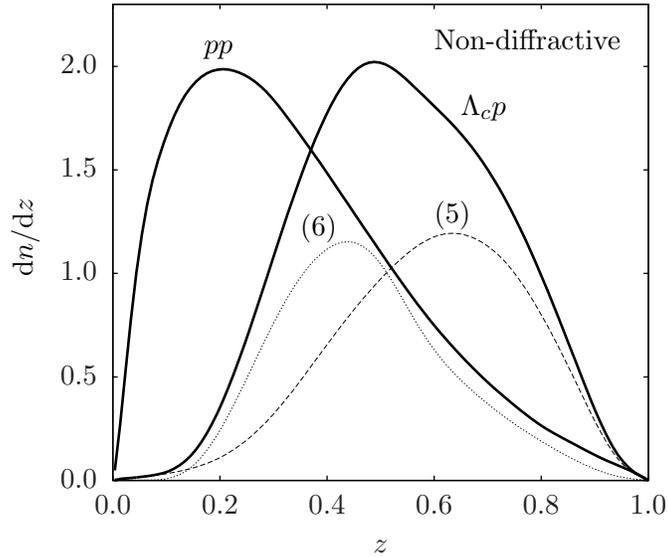} 
\end{center}
\caption{Fraction $z$ of energy taken by the leading baryon
or the charmed hadron in $p p$ and $\Lambda_c p$ 
non-diffractive collisions,
respectively. We have separated the distributions when the 
final state is a charmed baryon (5) or a charmed meson (6).
\label{fig3}}
\end{figure}

Let us first describe $\Lambda_c p$ non-diffractive collisions.
In Fig.~3 we plot the fraction $z$ of energy  carried by the
charmed hadron after the collision. We obtain that the leading
hadron may be a baryon (55\%) or a $D$ meson (45\%). The baryon
may be a $\Lambda_c^+$ or, with much smaller frequencies, 
$\Xi^0_c$,  $\Xi^+_c$ or  $\Omega_c^0$. The three meson species
($D^+,D^0, D_s$) appear with approximate 
frequencies of ($1:3:0.5)$. In Fig.~3
we have separated the spectra in the two cases. The average
fraction of energy taken by the baryons is $\langle z\rangle=0.60$
({\it i.e.}, $K_{\Lambda_c p}^{(5)} =0.40\pm 0.17$), whereas for the mesons
it is just $\langle z\rangle =0.47$ ($K_{\Lambda_c p}^{(6)}
=0.53\pm 0.15$). The events where the charm interacts with a light parton
in the target proton are mostly included among the ones with a
leading $D$ meson, and have an average elasticity of $\langle
z\rangle =0.42$. Finally, in 0.03\% of the partonic collisions the
incident charm anihilates (it is {\it traded} by a spectator sea
charm in the target proton) and $K_{\Lambda_c p}^{(7)} \approx 1$.
The average inelasticity of non-diffractive collisions is (we
include for comparison the inelasticity in $pp$ collisions)
\beq
K_{p p}^{n-di\! f}=0.66\,;\;\;\;\;
K_{\Lambda_c p}^{n-di\!f}=0.46\,.
\eeq

\begin{figure}
\begin{center}
\includegraphics[width=0.55\linewidth]{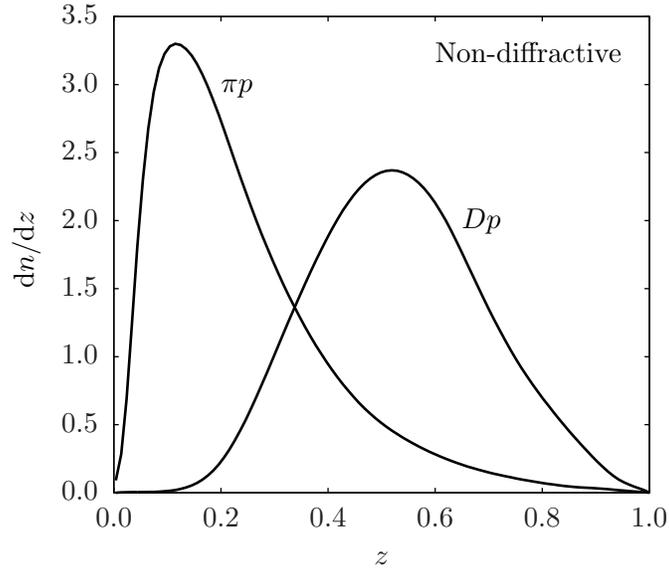} 
\end{center}
\caption{Fraction $z$ of energy taken by the leading baryon
or the charmed hadron in $\pi p$ and $D p$ 
non-diffractive collisions,
respectively. 
\label{fig4}}
\end{figure}

$Dp$ non-diffractive collisions are summarized in Fig.~4. The
final charm hadron is almost always (85\% of the events) a $D$
meson carrying a fraction $\langle z\rangle =0.55$ of the initial
energy ($K_{D p}^{(4)} =0.45\pm 0.16$). The $c$ quark anihilates ($K_{D
p}^{(5)} \approx 1$) with a sea $\bar c$ in the target just in 0.04\% of
the $D p$ partonic collisions. The inelasticity in these
non-diffractive processes is substantially lower than in pion
collisions, 
\beq 
K_{\pi p}^{n-di\! f}=0.77\,;\;\;\;\;
K_{D p}^{n-di\! f}=0.45\,. 
\eeq

\section{$H$--nucleus collisions}

\begin{figure}
\begin{center}
\includegraphics[width=0.55\linewidth]{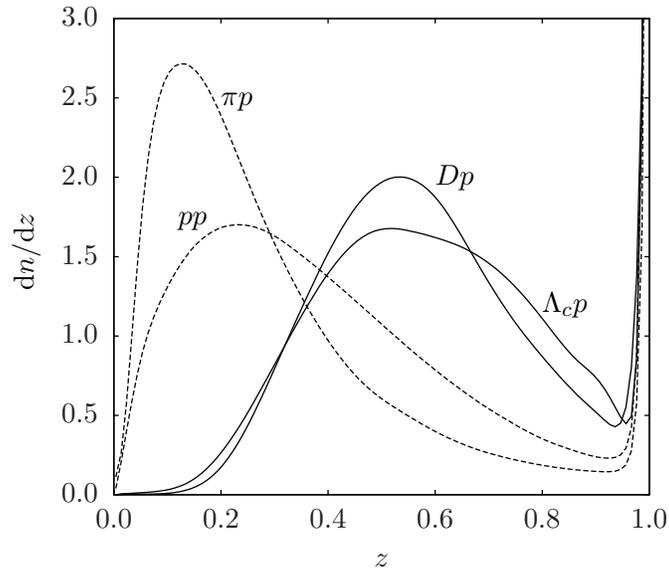} 
\end{center}
\caption{Fraction $z$ of energy taken by the leading baryon in 
$p p$ collisions, the leading pion in 
$\pi p$ collisions, 
and the charmed hadron in $\Lambda_c p$ and $D p$ 
collisions. The distributions include diffractive
and partonic collisions.
\label{fig5}}
\end{figure}
From the analysis in previous sections it results that the
inelastic cross section for the collision of a $10^9$ GeV charmed
hadron with a proton at rest is 
\beq 
\sigma^{\Lambda_c p}=88.4\;
{\rm mb}\,; \;\;\; \sigma^{D p}=59.2\; {\rm mb}\,, 
\eeq 
where
diffractive processes contribute a 30\% in $\Lambda_c p$
collisions and a 26\% in $D p$ interactions. 
The cross section for
proton and pion collisions 
of the same energy is around a 7\% larger, 
\beq
\sigma^{p p}=94.1\; {\rm mb}\,; \;\;\; 
\sigma^{\pi p}=64.8\; {\rm mb}\,. 
\eeq

The average
inelasticity in these collisions is then
\beq 
K_{\Lambda_c p}=0.39\,;\;\;\;\;
K_{D p}=0.38\,, 
\eeq 
which is substantially lower than the one 
in proton and pion collisions, 
\beq 
K_{p p}=0.59\,;\;\;\;\;K_{\pi p}=0.70\,. 
\eeq 
These results depend basically on the fragmentation model
used by PYTHIA, so they should not change substantially if the 
analysis
were based on the Montecarlo code SIBYLL \cite{Ahn:2009wx}
or on any other code using the Lund string scheme.

In this section we would like to comment on the approximate
effects of going from a proton to a nucleus target. The total
cross section for the collisions of $H$ with an atomic nucleus of
mass number $A$ can be estimated as 
\beq 
\sigma^{H A}\approx
A^{2/3} \sigma^{H p}\,, 
\eeq 
where the factor of $ A^{2/3}$ takes
into account the screening between the nucleons inside the
nucleus. For an averaged atmospheric nucleus of $A=14.6$ this
implies 
\beq 
\sigma^{\Lambda_c\, air}=528\; {\rm mb}\,; \;\;\;
\sigma^{D\, air}=354\; {\rm mb}\,. 
\eeq 
The associated interaction
length $\lambda^{H}_{int}=m_{air}/\sigma^{H\,air}$ in the
atmosphere when $H$ is a charmed meson or a baryon is therefore
\beq 
\lambda^{\Lambda_c}_{int}=46\; {\rm g/cm^2}\,; \;\;\;
\lambda^{D}_{int}=69\; {\rm g/cm^2}\,, 
\eeq 
which is a 6\% and a 9\% longer
than those of a pion and a proton of the same energy,
respectively.

To deduce the spectrum of the leading hadron after the collision
one must take into account that there may be more than one nucleon
involved in  each $H$--nucleus interaction. In general, it will be
possible to distinguish between {\it peripheral} and {\it central}
collisions. The first type is similar to the $H p$ processes
discussed before, and we will then assume that the spectra
coincide. Central collisions, on the other hand, imply a softer
spectrum of secondaries and a smaller value of the fraction $z$ of
energy taken by the leading charmed hadron.

As a first order estimate, we will assume that peripheral and
central collisions occur with equal frequency, and that the
average inelasticity in central processes is the typical in a
non-diffractive collision increased by $10\%$: 
\beq
K_{H\,air}\approx {1\over 2}\, K_{H\, air}^{peri}+ {1\over 2}\,
K_{H\, air}^{cent} \approx \left( {1\over 2}\, K_{H p} + {1\over
2}\, 1.1 \, K_{H p}^{n-di\! f}\right) \,. 
\eeq 
At $E=10^9$ GeV we obtain 
\beq 
K_{\Lambda_c\, air}\approx 0.45\,;\;\;\; K_{D\,
air}\approx 0.44\,. 
\eeq 
For proton and pion collisions the same
prescription gives an inelasticity 
\beq 
K _{p\, air}\approx
0.66\,;\;\;\; K_{\pi\, air}\approx 0.78\,.
\eeq 
This $12\%$ increase in $K$ when going from a proton to a
nucleus target compares well with the results obtained by 
other authors \cite{Ostapchenko:2009zz}. The frequency of
baryon to meson transitions in $\Lambda_c$--air central 
collisions would
be close to the 45\% obtained in partonic processes.

Finally, it is important to estimate the probability $p^{H\,
air}_{c\bar c}$ of a $c \bar c$ interaction that deposits all of
the charmed hadron energy. Short-distance interactions in nucleus
collisions scale proportional to the mass number, implying
\beq
p^{H\, air}_{c\bar c}\approx {A\sigma^{H p}_{c\bar c}\over A^{2/3}
\sigma^{H p}}\,.
\eeq
We obtain
\beq 
p^{\Lambda_c\, air}_{c\bar c}\approx 0.0015\,;\;\;\; 
p^{D\, air}_{c\bar c}\approx 0.0020\,.
\eeq

\section{Summary and discussion}

Charmed hadrons that decay weakly become {\it long lived} in the
atmosphere at energies $E> 10^8$ GeV, just like pions do at
$E>10^2$ GeV. The behaviour of pions in a calorimeter is well
understood and, as a consequence, their dynamics inside extensive
air showers is also known. We know, for example, that ultrahigh
energy pions are produced in these showers but never reach the
ground, since the atmosphere is {\it equivalent} to 10 meters of
water vertically and 30 times thicker from horizontal zenith
angles. In contrast, $D$ mesons are not directly observable at
calorimeters in colliders because they decay before they can reach
them. Any estimate of their possible effects in air showers must
then first {\it model} its hadronic interactions.

Here we have studied the collisions of charmed hadrons
with protons. Our intention has not been to perform an analysis
based on first principles, but to obtain an estimate of
the qualitative features of such processes based on PYTHIA.
The charm quark inside the hadron
carries a large fraction ($m_c/m_D\approx 0.7$) of
energy, so one may expect that the
spectrum of the final state will be substantially
different from the one in pion--proton collisions.
 We have argued that in most of the
processes either the charm can not be resolved ($q^2\ll m_c^2$)
or it is just an spectator. This allowed us to correlate them
with pion or proton collisions, and use then PYTHIA to simulate
quark fragmentation and the (strong or electromagnetic)
decay of higher-mass resonances.

We obtain that the average inelasticity in $D p$ collisions is
$K_{D\, p}=0.38$, a factor of 0.55 smaller than in $\pi p$
interactions. If the target is an air nucleus, we estimate values
around $K_{D\, air}=0.44$ and $K_{\pi\, air}=0.78$. These central
values would imply that starting with an initial $10^9$ GeV pion,
after 10 hadronic interactions in the atmosphere one is left 
with leading meson of just $300$
GeV. Instead, the energy of a $D$ meson would be
 reduced to $3\times 10^6$ GeV.

In contrast to pions and protons, we find that the collisions of
charmed mesons and baryons have a very similar inelasticity.
However, $\Lambda_c$ baryons have a large probability of becoming
$D$ mesons after a couple of interactions. This may actually be
an important source of $D$ mesons in extensive showers if the {\it
intrinsic charm} \cite{Brodsky:1980pb}
dominates the PDFs at high energy in proton collisions.

The proper inclusion of charm in extensive air-shower simulators
should describe both its production 
(already in DPMJET \cite{Berghaus:2007hp}) and also
its propagation. $D$ and $\Lambda_c$ hadronic collisions are
obviously irrelevant in colliders, which explains why they are
absent in codes like SIBYLL \cite{Ahn:2009wx}
 or QGSJET \cite{Ostapchenko:2007qb}. We think, however, that the
possibility to search for observable effects in astroparticle
experiments should make worthy an effort in such direction.

\section*{Acknowledgments}
We would like to thank 
Paolo Lipari, Davide Meloni and Sergio Sciutto
for discussions.
This work has been partially supported by
MICINN of Spain (FPA2010-16802, FPA2006-05294, 
and Consolider-Ingenio 
{\bf Multidark} CSD2009-00064) 
and by Junta de Andaluc\'{\i}a
(FQM 101 and FQM 437).

\end{document}